\RequirePackage{fix-cm}
\documentclass[twocolumn]{svjour3}          % twocolumn
\smartqed  % flush right qed marks, e.g. at end of proof
\usepackage{graphicx}
\usepackage{algorithm,algorithmic}
\usepackage{graphicx}
\usepackage{subfloat}
\usepackage{fancyhdr} 
\usepackage{amsmath}
\usepackage{fullpage}
\usepackage{amssymb}
\usepackage{amsfonts}
\usepackage{latexsym}
\usepackage{pstricks}
\usepackage{lineno}
\usepackage{lipsum}
\usepackage{afterpage}
\usepackage{authblk}
\usepackage{setspace}
\usepackage[toc,page]{appendix}
\usepackage{kantlipsum}
\usepackage{tikz}
\newcommand {\qq}  { {\sf q} }
\newcommand {\pp}  { {\sf p} }

\newcommand {\bfd}  { {\bf d} }

\newcommand {\kk}  { {\sf k} }
\newcommand {\DD}  { {D} }
\newcommand {\Dv}  { \sf{ Div} }
\newcommand {\Gr}  { \sf{ Grad} }

\newcommand {\Av}  { { A_{v}} }
\newcommand {\A}  { { A} }

\newcommand {\Sl}  { { S} }

\newcommand {\gb}  { {\sf g} }

\newcommand {\brho} { { \boldsymbol \rho} }
\newcommand {\bom} { { \boldsymbol \omega} }
\newcommand {\bsig} { { \boldsymbol \sigma} }
\newcommand {\bfepsilon} { { \boldsymbol \epsilon} }
\newcommand {\bfSigma} { { \boldsymbol \Sigma} }

\newcommand {\bfm}  { {\bf m} }
\newcommand {\bfu}  { {\bf u} }

\newcommand{\hf}{\frac12}

\newcommand{\g}{g}
\newcommand{\q}{q}
\newcommand{\sig}{\sigma}
\newcommand{\om}{\omega}

\newcommand{\diag}{{\sf diag}\, }

\journalname{Computational Geoscience}
\begin{document}

\title{Joint Hydrogeophysical Inversion: State Estimation  for Seawater Intrusion Models in 3D %\thanks{Grants or other notes
}

\titlerunning{Joint hydrogeophyscal inversion}        % if too long for running head

\author{Klara Steklova         \and         Eldad Haber %etc.
}

\authorrunning{Steklova, Haber} % if too long for running head

\institute{K. Steklova \at
              Earth and Ocean Sciences, The University of British Columbia, Vancouver, BC, Canada. \\
              \email{ksteklova@eos.ubc.ca}           %  \\
           \and
           E. Haber \at
              Earth and Ocean Sciences, \
							Mathematics, The University of British Columbia, Vancouver, BC, Canada.
}

\date{Submitted to Computational Geoscience, October 2015/ Revised, May 2016}

\maketitle

\begin{abstract}
Seawater intrusion (SWI) is a complex process, where 3D modeling is often necessary in order to monitor and manage the affected aquifers. Here, we present a synthetic study to test a joint hydrogeophysical inversion approach aimed at solving the inverse problem of estimating initial and current saltwater distribution. First, we use a 3D groundwater model for variable density flow based on discretized flow and solute mass balance equations. In addition to the groundwater model, a 3D geophysical model was developed for direct current resistivity imaging and inversion.
The objective function of the coupled problem consists of data misfit and regularization terms as well as a coupling term that relates groundwater and geophysical states. We present a novel approach to solve the inverse problem using an Alternating Direction Method of Multipliers (ADMM) to minimize this coupled objective function. The sensitivities are derived analytically for the discretized system of equations, which allows us to efficiently compute the gradients in the minimization procedure and reduce the computational complexity of the problem.
The method was tested on different synthetic scenarios with groundwater and geophysical data represented by solute mass fraction data and direct current resistivity data. With the ADMM approach, we were able to obtain better estimates for the solute distribution, compared to just considering each data set separately or solving with a simple coupled approach.

\end{abstract}

\section{Introduction}
\label{intro}
Seawater intrusion (SWI) is a complex process that occurs naturally due to small differences in density between freshwater and saltwater. Depending on the hydrogeological setting, seawater can enter the coastal aquifers through preferential flow pathways reaching far into the interior, or remain in close proximity to the coast \cite{Barlow09}. Increased groundwater (GW) extraction, reduced recharge into aquifers, and other human activities can cause the SWI to propagate further inland. To monitor the SWI and manage coastal aquifers, representative groundwater models need to be developed. Such models can then provide explanations for saltwater origin in the area, and can be used to validate different pumping scenarios to manage the saltwater front propagation and future freshwater demands \cite{Langevin12,Essink11,Kacimov09} and \cite{Sanford10}. 

For the affected areas we can usually expect some monitoring wells providing direct samples of fluid conductivity and thus indicating the salinity. However, understanding the complexity of SWI advance, this data might be too scarce to calibrate GW models and monitor the SWI progress. Geophysical methods offer an attractive option to map this process \cite{Nguyen09}. Saltwater in the pore spaces increases the soil bulk electrical conductivity, making it an ideal target for Direct Current (DC) resistivity or electromagnetic methods. Hence, these geophysical methods have become standard tools for monitoring SWI in the last two decades \cite{Bear99,Gottwein09,Trabelsi12,Nenna13,Fitterman14,Mills88} but have also many other environmental applications. Their popularity is mainly due to their ability to map the 3D subsurface non-intrusively and at a lower cost. 

One disadvantage of geophysical data is that they provide only indirect measurements; mapping the electrical conductivity of the subsurface which is dependent on other geological characteristics such as lithology rather than the amount of solute in groundwater alone. Furthermore, when collecting data, the choice of method and survey design has an effect on the depth resolution, and the area most sensitive to collected data. In contrast, groundwater well samples can provide the actual fluid conductivity, which in coastal areas is usually directly related to saltwater content. However, GW data may represent only a small space around the well or might not capture the difference between flow and resident concentration, see \cite{Carrera09} for more details. 

Combining both the geophysical and hydrological data offers  an attractive option to increase the amount of data for model calibration and to improve the estimates of GW states, which is also subject of this study. There are many different approaches in hydrogeophysical studies, often dependent on available data and software. In the uncoupled framework, geophysical and hydrogeological computations are independent. The advantage of this approach is that hydrogeological and geophysical models run independently, however, this also means that the a-priori information from the hydrology is not integrated into the geophysical inversion. Since geophysical inverse problems are ill-posed and require a regularization term, or a prior stochastic model (if Bayesian methods are used), ignoring hydrogeology data can severely deteriorate the quality of the geophysical inversion estimates. In contrast, in the coupled approach, the geophysical and groundwater models are linked together during the inversion. The hydrological state estimates are then guaranteed to be physically realistic and less a-priori assumptions are needed for the geophysical inversion. Due to this fact, coupled approaches were repeatedly found to give better results (\cite{Herck12}, \cite{Irving10} or \cite{Hinnell10}), however, at a cost of being computationally more intensive.

The computational cost is high also due to the GW flow governing equations, which for variable density flow, require solving a system of two strongly couple da nonlinear partial differential equations. Any 3D simulation is therefore costly when solving the inverse problem, where multiple GW forward simulations are necessary. To decrease the amount of estimated parameters and make the inverse problem solvable, usually some a-priori information based on geology is considered. For example by applying some geostatistical constraints when estimating the GW parameters \cite{Pollock12,Jardani13,Hermans12} or adding regularization terms to enhance smooth fields for estimated parameters or states.  The actual minimization is then often directed by some general inverse software as PEST \cite{Doherty11} or UCODE \cite{Poeter05} where the sensitivities are derived by perturbation approach. 

In the salt tracer experiments, the time and spatial scale enables to consider only solute transport, when modeling the GW flow, and thus simplify the GW inverse problem in terms of the computation cost. Examples can be found in work of Fowler and Moysey\cite{Fowler11}, where they investigated the non-uniqueness of GW parameter estimation using a single electrode and evaluating only the geophysical data misfit. A stochastic approach to coupled inversion was applied in Jardani et al.\cite{Jardani13} or Irving and Singha \cite{Irving10} by jointly inverting geophysical and groundwater concentration data during the tracer test to determine the hydraulic conductivity fields in 2D. In Monego et al. \cite{Monego10} estimated mean groundwater velocity and aquifer dispersivity (both assumed to be uniform) from the ERT mapped tracer test in a shallow heterogeneous aquifer.
 
For the SWI, the situation is usually more complex in terms of time scale and heterogeneity compared to solute tracer experiments. Many field studies were performed to delineate the SWI extent by use of geophysical data to indicate the saltwater intrusion progress \cite{Trabelsi12,Comte07,Nenna13,Fitterman14,Mills88}. Less work has been done in conjunction with GW data and estimating the GW parameters and states in the coupled approach.

In the work of Herckenrath et al.\cite{Herck12}, the authors compared two approaches for calibration of SWI model with the addition of TDEM data sets collected in Santa Cruz County, California. In the sequential approach the geophysical inversion for 1D TDEM soundings are run independently and geophysical estimates served as an extra observation for the GW model (after transforming via petrophysical relationship - Archie's law). Both data misfits were then minimized in the GW model inversion process. In the coupled approach the GW model is used to interpret the data and guide the geophysical inversion. Saltwater concentrations based on the 2D SEAWAT model \cite{Langevin08} were converted to electrical resistivity and 1D TDEM sounding forward responses were then calculated and compared with observed data. In total, six parameters were estimated for both the groundwater and TDEM model using the PEST optimization system. The authors concluded that the coupled approach provided a significant improvement in spatial resolution which would be hard to obtain with standard geophysical regularization techniques as was the case for the sequential approach.

A different way of transforming the information from geophysical estimates was introduced in Beaujean et al.\cite{Nguyen14}. The ERT derived conductivities were transformed via Archie's law to salt mass fraction estimates. These estimates were then filtered using a cumulative sensitivity based on squared Jacobians and served as extra data for hydrological inversion next to groundwater salt mass fraction data. The inversion was performed with PEST using a gradient based method.

The approaches above are somewhat limited to a small number of parameters because of the direct computations of sensitivities.
A more general framework was developed by Commer et al.~\cite{Commer14} that can be applied to larger scale hydrologeophysical problems for a wide range of processes in multiphase flow and solute transport. The authors improved the inversion framework of iTOUGH2 to enable parallel computing and merge it with parallel geophysical simulator for electromagnetic data. The sensitivities were evaluated by taking a perturbation approach. The high computational burden of this approach was balanced by the fact that the perturbed model simulation could be run independently.

While the work in \cite{Commer14} enables larger scale problems, it can be inefficient due to the finite difference evaluation of derivatives and hydrogeophysical coupling. Our goal here is to improve on this work such that we can solve large scale problems. To this end we present a new approach for the joint recovery of electrical conductivity and salinity. We solve the inverse problem by minimizing both types of data, where the petrophysical relationship provides a constraint in the minimization. The estimation of initial conditions for solute distribution is particularly important because if we have a good estimate of the current situation, we can better predict the future states, which is necessary for managing salt water intrusions.

We developed both geophysical and groundwater models, which are based on discretized systems of equations, in the same computational environment. Although similar models exist, we have developed our own flow and geophysical model to simplify the coupling implementation between physically different models. More importantly, it enables us to analytically derive the sensitivities necessary for solving the inverse problem. 
This allows us to deal with a large number of parameters at a cost that is cheaper than computing sensitivities using finite differences.

For the actual minimization of the coupled problem we have multiple options. We can use the knowledge of the petrophysical constraint and minimize both the data misfits simultaneously, for example by substituting the constraint and applying the Gauss Newton method for one variable only. This would result in a smaller problem, however in such case we have to deal with many regularization parameters to weight the different contributions to the coupled objective function by both data misfits.
Furthermore, since the relation between the different physical models is often empirical, forcing it may lead to artifacts (see \cite{Haber13} for further details).  
 
Therefore we introduce alternating direction method of multipliers (ADMM) to solve the coupled problem, which allows us to efficiently split the objective function into GW and geophysical parts and minimize each separately. The main advantage to this is that by separating the minimization of the coupled problem onto GW and geophysical parts only one regularization parameter is set during each iteration, and inversion codes stay relatively independent enabling for efficient parallelization.
Furthermore, since the petrophysics is enforced as a constraint, it is not applied exactly throughout the path
of optimization which yields additional degrees of freedom to the optimization algorithm.
 
ADMM has been introduced back in early 70's and has recently gained popularity for many inverse problems. It is a natural choice for multiphysics problems \cite{Boyd11}, also due to the strong convergence properties \cite{Ghadimi14}. It can be used in cases where the constraint (here the petrophysical relationship) can be considered as exact, in practice rather having low uncertainty.  A successful hydrological application can be found in Wohlberg et al. \cite{Wohlberg12}, where ADMM was applied to solve the inverse problem of estimating the piece-wise smooth hydraulic conductivity fields from sparse data for hydraulic head and conductivity, many more applications can be found in machine learning or statistical modeling.

The joint inversion scheme outlined above is expected to converge towards a solution that will fit both data sets. Difficulty converging may reveal discrepancies between the two sources of data, or between
the petrophysical model that is used to link the parameters. In our inverse problem setup with use synthetic groundwater and geophysical data (well salinity data end electrical potentials) generated only once. The unknown current and initial solute distribution are then estimated from this data. The fact that we solve for the solute content-electrical conductivity only, with the assumption of at least an approximate knowledge of other GW parameters can be regarded as naive, however, the same framework can be established for GW parameters such as hydraulic permeability, external fluxes and other GW variables as long as appropriate sensitivities are derived. We do not try here to estimate all of these parameters at once, since despite having two different sets of data, the inverse problem is essentially highly ill-posed, 
and most of these parameters vary over the entire domain. 
Our work is based on a frequentist approach to inverse problems \cite{tenorioScales} that is suited for large scale problems. Alternatively, if prior densities are known, one can use the Maximum A-Postriori estimate when considering the Bayesian framework and use similar techniques to the ones developed here. 

In the  Section~\ref{secGW} and \ref{secGeop}, we describe the groundwater model and geophysical model including the discretized form of governing equations. The authors are aware that these models are already well described in other literature but they are of a key importance in this study in order to follow the sensitivities derivation. In Section~\ref{IP}, we formulate the coupled inverse problem and introduced the new ADMM approach. The results for the joint inversion by ADMM versus a simple coupled approach are shown in Section~\ref{results}. We examined different synthetic cases of a pumping experiment in a coastal aquifer with homogeneous and heterogeneous permeability field and compared the joint approach (using ADMM) with a simpler coupled approach based on a reference model. Additionally, we looked at how the error of estimates changes if the GW model parameters used in the inversion deviate from the ``true'' parameters which served to create the synthetic data. Section~\ref{Disc} follows with discussion and finally, in Section~\ref{sum}, we summarize the paper and suggest future work.

%%%%%%%%%%%%%%%%%%%%%%%%%%%%%%%%%%%%%%%%%%%%%%%%%%%%%%%%%%%%%%%%%%%%%%%%%%%%%%%%%%%%%%%%%%%%%%%%%%%%%%%%%%%%%%%%%%%%%%%%%%%%%%%%%%%%%%%
\section{Groundwater model}
\label{secGW}
In this section, we briefly introduce the groundwater model used in this study, its discretization using finite volumes
and the solution of the discrete equations. Our method uses some specific properties of variable
density flow (VDF) to develop a highly efficient operator splitting method. 

\subsection{Governing equations}
The system of governing equations for VDF couples two processes: groundwater flow and solute transport, 
each represented by a partial differential equation in time and space.
The system can be written as:
\begin{gather}
\label{flow}
\nabla \cdot (\rho \q) = \rho Q_{gw}\\
\label{solute}
\frac{ \partial (\phi~\rho\omega)}{\partial t} + \nabla \cdot(\phi~\rho \DD \nabla \omega) - \nabla \cdot (\rho  \omega~ \q) = Q_{s\omega}
\end{gather}
where
\begin{equation}
\label{velocity}
\q(\rho) = -\frac{k}{\mu}(\nabla p - \rho \g \nabla z). 
\end{equation}

The system above is a pressure  - solute mass fraction formulation \cite{Diersch98}, where $\rho$ is the fluid density, $\omega$ is the  fraction of solute (saltwater) in the fluid,  $\phi$ porosity, $\DD$ represents the hydrodynamic dispersion and $Q_{s\omega}$ and $Q_{gw}$ the external fluxes of solute and groundwater and $t$ stands for time.
 $q$ is the groundwater velocity based on Darcy's law \cite{Bear99}, where $p$ is the pressure, $k$ the permeability of porous media, $\mu$ the fluid viscosity, $\g$ the gravitational constant and $z$ is a downward coordinate direction.
To complete the system a number of physical relationships and parameters are required.
We assume that the mass fraction $\omega$ is connected to the density by the linear relation
\begin{equation}
\label{rho}
\rho = \rho_{F}(1+ \gamma \omega) ~~\mbox{with~~}\gamma = \frac{\rho_{S} - \rho_{F}}{\rho_{F}},
\end{equation}
where $\rho$ represents the fluid density, $\rho_{F}$ freshwater density and $\rho_{S}$ saltwater density.

In our model we used a number of simplifications that can be relaxed, but are justified for our application. 
First, we assume a steady state for groundwater flow equation \eqref{flow}, which still has to be resolved throughout each time step computation to update the pressure and velocity field as a result of solute content dynamics. Next, the hydrodynamic dispersion tensor $D$ is kept fixed in our model. Using a full variable density flow model, as for example in \cite{Diersch98}, would not change the approaches described in the following sections regarding coupling and solving the inverse problem.

\subsection{Discretization and solution of the groundwater model}
\label{GW_num}
In this subsection, we briefly discuss the discretization used for each of the governing equations. Even though the following text is not necessary to understand the idea behind the hydrogeophysical coupled approach in section \ref{IP}, it is an important part in our study, since we derive the sensitivities analytically based on the discretized equations, which are later needed for the Gauss-Newton method.
We use a cell-centered finite volume method for the flow equation \eqref{flow}, and an operator splitting method for the solute transport equation \eqref{solute}.
 In particular, when integrating the solute transport we use a Semi - Lagrangian
method for the advection part and add the dispersion part implicitly.

\subsubsection{Fluid mass balance equation}

Our groundwater model discretizes equations~\eqref{flow} and \eqref{solute} in 3D on a staggered grid.
The  solute fraction $\omega$,  and the pressure $p$ are placed in the cell centers and the fluxes, $\q$, on
the cell faces.
The parameters $\phi$ and $\DD$ are discretized at the cell centers and harmonically averaged onto cell faces when needed. For a complete description on the discretization of systems of the form 
\eqref{flow} and \eqref{velocity} using finite volumes see \cite{Haber14}. The discrete pressure equation reads
\begin{equation}
\label{flowdis}
\Dv~R_{KM}  \left(\Gr~ \pp^{n} + \left(\Av \brho^{n}\right) \odot \gb \right) = \brho^{n}\odot Q_{gw}
\end{equation}
where $R_{KM} = \diag(\Av(\frac{\boldsymbol \mu}{\brho \kk}))^{-1}$ and the division is done pointwise, $\Dv,\Gr$ are divergence and gradient matrix operators. $\Av$ is the arithmetic averaging operator and $\gb  = - g~ \Gr~{\sf z}$ is a gravity acceleration vector. We use the Hadamard product ${\sf a} \odot {\sf b}$
for the element wise product of two vectors. The ad-script $n$ stands for the corresponding hydrological state at the $n^{th}$ time step.  

Assuming that we know the density at the $n^{th}$ time step, the unknown pressure $\pp^{n}$ can be solved directly
\begin{eqnarray} 
\pp^{n} = && (-\Dv~ R_{KM} \Gr)^{-1}\nonumber\\
          && (\Dv~R_{KM}~\left(\Av \brho^{n}\right) \odot \gb  -\brho^{n}\odot Q_{gw}),
\end{eqnarray}
where the matrix $\Dv~ R_{KM} \Gr$ is inverted using either Cholesky (for small to medium scale problems) or by the preconditioned conjugate gradient method.
Given the pressure $\pp^{n}$ we can compute the groundwater linear velocity $\qq^{n}$  at the cell faces:
\begin{equation}
 \qq^{n} = - \diag\left(\frac{1}{\Av(\frac{{\boldsymbol \mu}{\boldsymbol \phi}}{\kk})}\right)\left(\Gr~\pp^{n} +  \left(\Av \brho^{n}\right) \odot \gb\right).
\end{equation}

\subsubsection{Solute mass balance equation}
  For the solution of equation \eqref{solute}, we use operator splitting.  The system is split into an advection and a dispersion part
 and then integrated sequentially from time $t_{n}$ to $t_{n+1}$:
 \begin{subequations}
 \begin{eqnarray}
 \label{split1}
\frac{ \partial (\phi~\rho\omega)}{\partial t} - \nabla \cdot (\rho  \omega~ \q) &=& 0 \\
 \label{split2}
 \frac{ \partial (\phi~\rho\omega)}{\partial t} + \nabla \cdot(\phi~\rho \DD \nabla \omega) &=& Q_{s\omega} 
\end{eqnarray}
\end{subequations}

In operator splitting methods, the advection equation is typically solved first, using some explicit method starting at $\omega^{n}$ to obtain the temporary variable $\omega^{*}$. The diffusion equation is solved next using an implicit method and obtaining $\omega^{n+1}$ from $\omega^{*}$.

For the problems in this work, the advection equation is solved using a Semi-Lagrangian approach and dispersion is solved with an Eulerian step. Our method belongs to the family of modified methods of characteristics (MMOC) first introduced in \cite{Russell82}. The main advantages are in the alleviation of the Courant number restriction due to the Lagrangian advection step \cite{Celia90} and mass conservation. The Eulerian - Lagrangian scheme is also effective in overcoming numerical dispersion for advection dominated problems \cite{Sorek11}. 

Similar approaches are for example taken by codes such as MOC3D model for solute transport \cite{Konikow96}, or later MT3DMS for variable density flow in connection with Modflow \cite{Langevin06,Langevin08}. In the context of review on Eulerian-Lagrangian localized adjoint methods (ELLAM)\cite{Russell02} our approach is a finite difference Eulerian-Lagrangian type, where we do not solve the solute transport equation using an integral equation but with a projection matrix.

\begin{figure}
\label{picFig}
\begin{center}
\begin{tikzpicture}
%    \fill[blue!20] (1.1,2.35) -- (1.1,1.35) --  (2.2,0.35) -- (3.1,3.35) -- cycle;
  \filldraw [red] (0.65,0.65) circle (5pt);
   \draw [->,black, thick](0.625,0.625) to (2.7,2.4);
  \filldraw [black] (2.8,2.5) circle (5pt);
  
  \filldraw [gray] (3.25,3.25) circle (5pt);
  \filldraw [gray] (1.95,3.25) circle (5pt);
  \filldraw [gray] (1.95,1.95) circle (5pt);

  \filldraw [gray] (3.25,1.95) circle (5pt);

\draw [->,green, thick](2.7,2.4) to (3.25,3.25);
\draw [->,green, thick](2.7,2.4) to (1.95,3.25);
\draw [->,green, thick](2.7,2.4) to (1.95,1.95);
\draw [->,green, thick](2.7,2.4) to (3.25,1.95);
  {\draw[step=1.3cm] (-1,-1) grid (5,5);}
   
\end{tikzpicture}
\caption{Particle and cell discretization of the transport equation \label{discAdvec}: The exact trajectory of the particle is followed based on the known velocity, and the mass of the particle is the distributed among neighboring cell centers by linear interpolation.}
\label{pic_SL}
\end{center}
\end{figure}
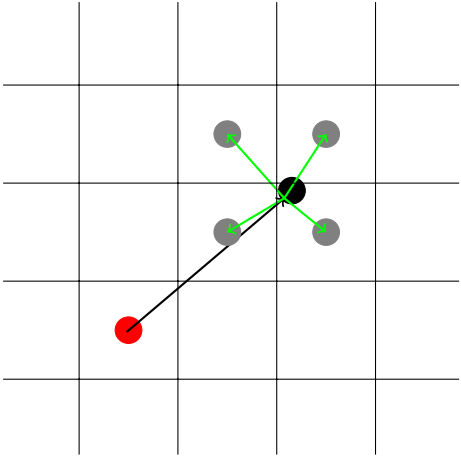

In our implementation, particles are placed at the mesh centers at each time step and tracked  forward.
The mass of each transported particle is interpolated to its neighbors yielding the solution at the next time step (Figure~\ref{pic_SL}).  

Rather than transporting solute mass fraction $\omega$, we transport solute mass $\omega \rho$
at each time step, and since the Lagrangian method is conservative, $\omega \rho$
is conserved. The discretization of the advection equation can be written as
\begin{eqnarray}
\label{discAdvec}
\centering
(\brho \odot \bom)^{*} = \Sl^{n} (\brho \odot \bom)^{n}
\end{eqnarray}
where $ \Sl^{n}$ is an interpolation matrix that contains the interpolation weights and is a function of the velocity field $\qq^{n}$.
Using the condition \eqref{rho}, we can solve a local quadratic equation for $\bom^{*}$, noting that
only one root of the equation makes physical sense.

Having the solution  $\bom^{*}$, we can now integrate the diffusion dominated part, starting
from $\bom^{*}$. With an implicit Euler method, the equation reads
\begin{eqnarray}
\label{diffimp}
&&\frac{(\brho\odot\bom)^{n+1} - \Sl^{n} (\brho\odot\bom)^{n}}{\Delta t}  = \nonumber\\ 
&&\Dv~ {\rm diag}\left(\Av {\frac{1}{\DD \odot \brho(\bom^{n+1})}} \right)^{-1} ~\Gr~\bom^{n+1} + Q_{s\omega}. \nonumber
\end{eqnarray}
The implicit diffusion is nonlinear and it is solved using a Picard iteration \cite{Ackerer04}, updating also the velocity field and corresponding $\bom^{*}$.
For VDF with seawater, the density difference between the two fluids is fairly small, decreasing the nonlinearity of the coupled system and the number of Picard iterations. Even though the time step can be large given the stability of the Semi - Lagrangian method, care must be taken with respect to the coupling with the flow equation \cite{Putti95}. A step too large would lead to a weak coupling and the possibility of erroneous calculations.

\subsection{Groundwater model sensitivities  \label{GW-sens}}
In order to proceed with Gauss - Newton minimization when solving the inverse problem, we need to know the sensitivities of the collected data with respect to the unknown initial solute content. In our inverse problem, the data are represented by the final solute mass fraction. At each time step of the groundwater forward model simulation we solve the system of two partial differential equations, where at the end of each time step the pressure $\pp^{n}$ is given by the solute distribution and boundary conditions. The groundwater velocity $\qq^{n}$ can be then expressed as a function of $\bom^{n}$ and $\pp^{n}$ only and the system reduces to the second equation for solute transport. The solute mass balance equation \eqref{solute} is time dependent, and the velocity dependence is stored in interpolation matrix $S^{n}$, which is a function of flux $\qq^{n}$ and solute mass $(\brho \bom)^{n}$. \\
Since the sensitivities do not need to be known exactly, we can assume that for their calculation the density does not vary much and proceed with using the solute mass fraction formulation alone, for which the time stepping process of equation \eqref{solute} can be written as:
\begin{equation}
 (I - \Delta t M_{dis})\bom^{n} - S^{n-1}\bom^{n-1} - \Delta t~q_{BC, ex}^{n-1} = 0
\end{equation}
Since we are using the operator splitting approach, we can explicitly derive the sensitivity of the mass $\bom^{n+1}$ at each time step with respect to the solute fraction at the previous time step. This is a two step calculation, step 1 being the advection part and step 2 being the diffusion part. The sensitivity of the final solute fraction can therefore be calculated recursively during the forward GW model run.

\begin{figure*}
\includegraphics[width=0.95\textwidth]{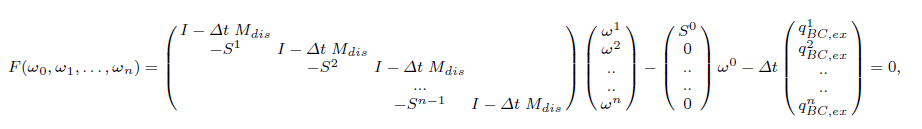}
\caption{\small{Full forward GW model for all time step at once.}}
\label{eqinpic}
\end{figure*}

More generally, if we consider all time steps together (see in Figure \ref{eqinpic}),
 which can be written in a compact form as
\begin{eqnarray}
\label{ttimeSteppingCond}
A_{TS} \bom - B_{0}\bom_{0} - \widehat q = 0,
\end{eqnarray}
where $A_{TS}$ is the time stepping matrix that is block bidiagonal and $B_{0}$ is the matrix
that multiplies $\bom_{0}$. The vector $\widehat q$ involves the boundary conditions
and sources 
and vector $\bom$ is set here as $\bom = [\bom_{1}^{\top},\ldots,\bom_{n}^{\top}]^{\top}$.	

Differentiating $F$ with respect to $\bom_{0}$ we obtain that
\begin{equation} {\frac {\partial F(\bom,\bom_{0})}{\partial \bom}} \frac{\partial \bom}{\partial \bom_{0}} + {\frac {\partial F(\bom,\bom_{0})}{\partial \bom_{0}}} = 0,
\end{equation}
and therefore
\begin{equation}
{\frac {\partial \bom}{\partial \bom_{0}}} = - {\frac {\partial F(\bom,\bom_{0})}{\partial \bom}}^{-1} {\frac {\partial F(\bom,\bom_{0})}{\partial \bom_{0}}}. 
\end{equation}
Using \eqref{ttimeSteppingCond} we see that
\begin{eqnarray}
\label{sensw}
{\frac {\partial \bom}{\partial \bom_{0}}}  = A_{TS}^{-1}B_{0} 
\end{eqnarray}	
and later on, this is referred to as the sensitivity $J$.
In particular, if we wish to compute the sensitivity of data that are measured for $\bom$ in a small number of locations/times
with respect to $\bom_{0}$, we extract these  points from \eqref{sensw}.  Let the matrix $Q_{f}$ be the matrix that extracts
$\bom$ for the time steps and locations that the data are measured in
\begin{equation} \bfd_{f} = Q_{f} \bom.
\end{equation}
Then, the sensitivity of the last time step is simply
\begin{eqnarray}
\label{sensnw}
{\frac {\partial \bfd_{f}}{\partial \bom_{0}}}  = Q_{f}A_{TS}^{-1}B_{0}.
\end{eqnarray}	

It is important to note that in order to compute the sensitivity times a vector, one does not need to compute
the sensitivity explicitly in a matrix form. Instead one needs to multiply $B_{0}$ times a vector and then solve a time stepping
problem forward. For the multiplication of $\left({\frac {\partial \bfd_{f}}{\partial \bom_{0}}} \right)^{\top}$
times a vector, one multiplies by $Q_{f}^{\top}$ and then solves a backward time stepping process.
For more details about the implementation see \cite{Haber14}.

%%%%%%%%%%%%%%%%%%%%%%%%%%%%%%%%%%%%%%%%%%%%%%%%%%%%%%%%%%%%%%%%%%%%%%%%%%%%%%%%%%%%%%%%%%%%%%%%%%%%%%%%%%%%%%%%%%%%%%%%%%%%%%%%%%
%%%%%%%%%%%%%%%%%%%%%%%%%%%%%%%%%%%%%%%%%%%%%%%%%%%%%%%%%%%%%%%%%%%%%%%%%%%%%%%%%%%%%%%%%%%%%%%%%%%%%%%%%%%%%%%%%%%%%%%%
\section{Geophysical imaging method}
\label{secGeop}
In order to estimate the physical properties of the media, we choose direct current (DC) resistivity, also referred to as Electrical Resistance Tomography (ERT). DC resistivity is sensitive to the electrical conductivity of the media and since the conductivity of salt water is a few orders of magnitude larger than that of freshwater, the DC resistivity method is a natural choice. Other electrical methods can be used, however, DC has several advantages, i.e. data acquisition is relatively easy, computational effort is minimal, and cost is low.

In DC experiments, a current is injected into the ground  creating an electrical potential, which is then measured using pairs of electrodes placed on or under the surface. To model this process, we used the steady state form of Maxwell's electromagnetic equations: \begin{gather}
\nabla \cdot (-\sigma \nabla \varphi) = \textup{\bf{I}}(\delta(\bf{r}-\bf{r}_{s^+})  - \delta(\bf{r}-\bf{r}_{s^-}))\\
{\bf n}\cdot \nabla \varphi = 0 ~~\mbox{on}~~ \Gamma_{nc} \nonumber
\end{gather}
where $\sigma$ represents the media's electrical conductivity, $\varphi$ is the electric potential, $\bf I$ is the current source, $\delta$  Dirac delta function and $r_{s+,s-}$ stands for location of positive and negative electrodes. The boundary conditions were set as no flux across the boundaries, $\Gamma_{nc}$. When solving the forward problem, the electrical conductivity field, $\sigma$, is known and potentials everywhere can be calculated using a finite volume approach on a 3D grid. 
Since the discretized DC equation is essentially the same type of equation (Poisson equation) as the flow equation (\ref{flowdis}), we follow the same procedure to solve it as has already been described in the previous section and use a cell-centered finite volume approach for the discretization of the problem leading to a linear system of equations: \begin{equation}
 \A(\bsig)\bfu =  \qq
\label{Agphy}
\end{equation}
where \begin{equation}
\A(\bsig) = - \Dv~\diag\left( \frac{1}{(\Av \bsig^{-1})}\right) \Gr. 
\end{equation}
The electrical conductivity $\bsig$ is averaged harmonically from the cell centers onto cell faces, $\qq$ is the source term and $\bfu$ represents the potentials.
The forward model solves the potential values everywhere for a given conductivity field, and using the data projection matrix $Q_{e}$, the observed data are
\begin{equation}
\label{edata}
{\bf d}_{e} = {Q_{e}} \bfu,
\end{equation}
 measured at the receivers.

\subsection{Sensitivities}
\label{DC_sens}
We can write the forward geophysical model $G(\bfm,\bfu)$ in simple matrix vector notation as: 
\begin{equation} 
G(\bfm,\bfu) = A(\bfm) \bfu - \qq = 0,
\end{equation}
where $\bfm$ is commonly chosen to be the log electrical conductivity, i.e. $\bfm = log(\bsig)$  and following Eq.\eqref{Agphy} $A(\bfm)$ is just
\begin{equation}
A(\bfm) =  - \Dv~\diag\left( (\Av exp(\bfm)^{-1})^{-1}\right) \Gr 
\end{equation}
Following the basic rules of sensitivity calculation, the derivative of the forward geophysical model with respect to $\bfm$ is equal to zero:
\begin{equation}
\frac{\partial G(\bfm,\bfu) }{\partial \bfm} + \frac{\partial G(\bfm,\bfu) }{\partial \bfu}\frac{\partial \bfu}{\partial \bfm} = 0.
\end{equation}
The sensitivity of potential $\bfu$ with respect to $\bfm$ is then 
\begin{eqnarray}
\frac{\partial \bfu}{\partial \bfm} &&= - \left(\frac{\partial G(\bfm,\bfu) }{\partial \bfu}\right)^{-1}\frac{\partial G(\bfm,\bfu)}{\partial \bfm}\nonumber\\
&& = - A(\bfm)^{-1}\frac{\partial G(\bfm,\bfu)}{\partial \bfm}.\end{eqnarray}
We can then substitute into \eqref{edata} and obtain the sensitivity of measured data with respect to log conductivity $\bfm$ as
\begin{equation}
\frac{\partial {\bf d}_{e}}{\partial \bfm} = {Q_{e}} \frac{\partial \bfu}{\partial \bfm}.
\end{equation}
The matrix $\frac{\partial G(\bfm,\bfu)}{\partial \bfm}$ is a sparse matrix and its calculation is discussed in
\cite{PidliseckyHaberKnight2006b}. Again, the sensitivity matrix does not need to be evaluated explicitly, only matrix vector products are used in the Gauss-Newton minimization procedure.

\subsection{Salt mass fraction and electrical conductivity relationship}
The electrical conductivity of a porous media is dependent on fluid conductivity, rock/soil porosity, permeability,  saturation, temperature and also mineral composition \cite{Kemna_book}. In the saturated sediments of coastal aquifers, fluid conductivity is usually the main factor due to salinity of the fluid, but in other cases surface conductivity of fluid-grain interface, porosity or amount of saturation can be the main factors influencing the overall electrical conductivity. Empirical petrophysical relationships, such as Archie's law \cite{Archie47}, were therefore developed  to relate these properties. For fully saturated sediments Archie's law can be written as:
\begin{equation} \sig_{b} = \frac{1}{\alpha}\sig_{w} \phi^{n} \end{equation}
where $\sig_{b}$ is the bulk electrical conductivity, $\sig_{w}$ is the fluid electrical conductivity, $\phi$ porosity. $\alpha$ and $n$ are empirical parameters related to rock type which can be calibrated based on soil core samples, field survey or estimated by recommended values based on geological characterization. 

The fluid electrical conductivity in natural waters increases with the amount of dissolved solids and ions \cite{Hem85}. Linear relationships can be found for a fixed temperature between total dissolved solids (TDS) and fluid electrical conductivity.
Since the seawater is usually dominated by sodium chloride ions, we can assume a linear relationship between salt mass fraction and fluid conductivity: 
\begin{equation}
\sigma_{w} = c~ \om + \sig_{F},
\end{equation}where $\sig_{F}$ is the conductivity of freshwater, $c$ a constant and $\om$ is the salt mass fraction. The electrical conductivity of seawater is approximately two orders of magnitude higher than that of freshwater. After substituting into Archie's law, we have
\begin{equation}
\label{MyArchie}
\sigma_{b} =  \frac{1}{\alpha}(c \omega + \sig_{F}) \phi^{n}.
\end{equation}
In this petrophysical relationship one assumes that the bulk conductivity is affected only by the electrical conductivity of the fluid in the porous matrix, the variations in temperature are small, and that the surface conductivity of porous material is negligible. This would not be valid for example, in environments with a high clay content \cite{Kemna_book}; in many cases though, the differences in conductivity due to the variation in geological material are negligible compared to the increase in conductivity due to saltwater ions.

Equation (\ref{MyArchie}) was used to generate electrical conductivity models based on the salt mass fraction from groundwater model simulations.

In this study we refer to groundwater data as solute mass fraction data, $\omega$, even though in the field applications it is more common to record the fluid conductivity $\sig_{w}$ of GW samples, and only later by using a linear relationship to transform it to an actual solute mass fraction. However, this linear relationship is different from the general petrophysical relationship, having a different error and lower uncertainty. Also, the groundwater model is defined for solute mass fraction, and this way it is easier to distinguish in the following context the groundwater and geophysical origin of data.
%%%%%%%%%%%%%%%%%%%%%%%%%%%%%%%%%%%%%%%%%%%%%%%%%%%%%%%%%%%%%%%%%%%%%%%%%%%%%%%%%%%%%%%%%%%%%%%%%%%%%%%%%%%%%%%%%%%%%%%%%%%%%%%%%%%%%%%%%%%%%%%%%%%%%%%%%%%%%%%%%%%%%%%%%%%%%%%%%%%%%%%%%%%%%%%%%%%%%%%%%%

\section{Solving the inverse problem}
\label{IP}
Assume now that we have obtained DC resistivity and groundwater well samples data at a single time, that is
we have
\begin{eqnarray}
\label{data}
\bfd_{e} &=& Q_{e} \bfu + \bfepsilon_{e} \\
\bfd_{f} &=& Q_{f} \bom + \bfepsilon_{f} 
\end{eqnarray}
where $Q_{e}$ and $Q_{f}$ are projection matrices that project the electrical potential field, $\bfu$, and the solute fraction field, $\bom$ onto their measurement locations, respectively. Let $\bfepsilon_{e}$ and $\bfepsilon_{f}$ be vectors with the errors associated with each measurement, which are assumed to be Gaussian, independent and identically distributed with covariance matrices $\bfSigma_{e}$ and $\bfSigma_{f}$. The $\bom$ represents the solute fraction at the end of simulation, i.e. at the same time as the geophysical data were collected. 

The goal is to jointly invert the different data in order to better recover the flow path and in 
particular, to better predict the flow. Clearly, the electric data is affected mainly by the electrical conductivity of the porous media in SWI and the groundwater data is
affected by the initial solute distribution, $\omega_{0}$,  the porosity, $\phi$, the dispersion, $D$, the permeability
$k$ and the fluid viscosity, $\mu$. If all GW model parameters are unknown, then the indirect data or scarce direct data
will not suffice to accurately estimate them. However, 
well studied aquifers are typically observed for many years and
therefore, for now we assume that all parameters, other than solute fraction, are at least approximately known. For example some of these parameters could be estimated by first solving the GW inverse problem using pressure or hydraulic head data, which are not considered in the coupled problem here. Thus we write $\bom(\bom_{0})$ and solve for the initial conditions only, and to also recover the current solute fraction distribution.
\bigskip

We now develop a procedure to estimate the electrical conductivity $\bsig$ and the initial
solute fraction distribution $\bom_{0}$ by  the regularized maximum likelihood  estimate \cite{Haber14,Medina96,Doug04}. This leads to the following constrained optimization
problem
\begin{eqnarray}
\label{min1}
\min_{\bsig,\bom_{0}} && \alpha_{e} \hf \|{\bf d_{e}} - Q_{e} \bfu(\bsig)\|^{2}_{\bfSigma_{e}^{-1}}+ \beta_{e}R(\bsig)  \nonumber\\
									   && +  \alpha_{f} \hf\|\bfd_{f} - Q_{f} \bom(\bom_{0})\|^{2}_{\bfSigma_{f}^{-1}} + R(\bom_{0}) \\
\label{minst}
{\rm s.t} && \bsig = \eta\bom + \bsig_{fb} = p(\bom) \nonumber
 \end{eqnarray}

 Here $R$ is a regularization operator (to be discussed next) and $\alpha_{e},\alpha_{f}$ and $\beta_{e}$ are regularization
 parameters. The constraint \eqref{minst} represents Archie's law and is obtained from equation \eqref{MyArchie} by lumping
 a few parameters into $\eta = \frac{c}{\alpha}\phi^{n}$, and $\bsig_{fb}$ being the background soil bulk conductivity, which is equal to $\bsig_{fb} = \frac{1}{\alpha}\phi^{n}\sigma_{F}$ and corresponds to to the conductivity for porous media with freshwater only.

A number of different approaches may be taken to solve this constrained optimization problem outlined in Equation \eqref{min1}. One possibility is to eliminate $\bsig$ and to work with $\bom_{0}$ alone. The objective function (where we omit the extra regularization on $\bsig$) is then 
\begin{eqnarray}
\label{min3}
\Phi(\bom_{0}) =  \alpha_{e}\frac 12 \|{\bf d_{e}} - Q_{e} \bfu\left(\bsig\left(\bom(\bom_{0})\right)\right)\|^{2}_{\bfSigma_{e}^{-1}} \nonumber\\
+ \alpha_{f}\hf \|\bfd_{f} - Q_{f} \bom(\bom_{0})\|^{2}_{\bfSigma_{f}^{-1}} + R(\bom_{0}) 
 \end{eqnarray}
 This approach has the advantage of solving a smaller problem, however, it complicates other aspects of the inversion. First, we are required to choose two regularization parameters, $\alpha_{e}$ and $\alpha_{f}$ for the different data misfits with respect to the regularization term at each iteration. Despite abundant research on choosing one regularization parameter, there are almost no criteria for setting two parameters during the iterative minimization process where each data misfit has a different speed of convergence. 
 A more detailed discussion can be found in \cite{Commer09} where the authors used magnetotelluric data together with controlled source EM data. Even though both data sets can be modeled by changes  in conductivity, non-trivial weighting was needed in order to jointly invert them. 
Second, the multiplication of Jacobians and data misfit calculations can have very different computational cost for each data misfit, which can make each iteration very unbalanced and does not favor parallelization.
Finally, if the relationship between $\bsig$ and $\bom_{0}$ are inexact, forcing them may lead to inversion 
artifacts.

Therefore, we opt to use the alternating direction method of multipliers (ADMM) to minimize this coupled objective function. The main advantage of ADMM is that the GW and geophysical parts can be solved separately, that is, we do not need to weight the two different data misfits in one objective function, but instead we split the minimization into two subproblems. This enables using existing inversion methodologies and
even software packages only with minor changes. It also takes into advantage of the existing parallelization for a single problem and therefore substantially increase the efficiency when solving  large scale problems.  
 
Following the ADMM approach, the augmented Lagrangian for \eqref{min1} is
 \begin{eqnarray}
 \label{auglag}
{\cal L}(\bsig,\bom,{\bf y}) = && \frac 12 \|\bfd_{e} - Q_{e} \bfu(\bsig)\|^{2}_{\bfSigma_{e}^{-1}} + \beta_{e}R(\bsig) \nonumber\\
+&&\hf \|\bfd_{f} - Q_{f} \bom(\bom_{0})\|^{2}_{\bfSigma_{f}^{-1}} + \beta_{f}R(\bom_{0}) \nonumber\\ 
+&&{\bf y}^{\top}(\bsig - p(\bom)) + {\frac {\rho}2} \|\bsig -  p(\bom ) \|^{2}
\end{eqnarray}
Here ${\bf y}$ is a Lagrange multiplier and $\rho$ is a parameter that can be chosen somewhat arbitrarily. 
The $k^{th}$ iteration of  ADMM is summarized in Algorithm~\ref{alg1}. 
 \begin{algorithm}
\caption{ADMM}
\label{alg1}
   \begin{algorithmic}
         \STATE $\bullet$ Approximately minimize the augmented Lagrangian with respect to $\bsig$.
         \STATE $\bullet$ Approximately minimize the augmented Lagrangian with respect to $\bom_{0}$.
	 \STATE $\bullet$ Update the Lagrange multiplier ${\bf y}^{k+1} = {\bf y}^{k} +
	\rho (\bsig^{k+1} -  p(\bom^{k+1}))$.
	 \end{algorithmic}
\end{algorithm}

We now discuss the solution of each subproblem and show that by using small modifications to existing
inversion codes, the ADMM iteration can be carried out efficiently.\\

\subsection{Geophysical imaging block descent}
At each step of minimizing the augmented Lagrangian with respect to $\bsig$, we approximately solve:
\begin{eqnarray}
 \label{gmin}
\min_{\bsig} \Phi(\bsig) = && \frac 12 \|\bfd_{e} - Q_{e} \bfu(\bsig)\|^{2}_{\bfSigma_{e}^{-1}} + \beta_{e}R(\bsig) \nonumber \\
												   && +{\bf y}^{\top}(\bsig -  p(\bom )) + {\frac {\rho}2} \|\bsig -  p(\bom ) \|^{2} 
\end{eqnarray}
 The objective function consists of a data misfit and a regularization part, as in usual inverse problems, and coupling terms involving also $\bom$, the groundwater variable.  We can proceed using the Gauss-Newton method to minimize \eqref{gmin} with respect to $\bsig$. Compared to a standard inverse problem, we need to know also the derivatives of the coupling terms with respect to $\bsig$, however, these are, in this case, straightforward since $\bom$ is fixed.

Here, we consider a quadratic regularization of the form 
\begin{equation}R(\bsig) = \frac 12 \beta_{e} \left\|L (\bsig - \bsig_{\rm ref})\right\|^{2},\end{equation}
 where $L$ is the gradient operator and $\bsig_{\rm ref}$ can be either set to the background conductivity reference model, or derived using Archie's law from the last estimate via the groundwater descent step.
If we set $J_{e} = \frac{\partial \bfu(\bsig)}{\partial \bsig}$ and assume for simplicity that $\Sigma_{e} = I$, the derivative $\frac{\partial \Phi }{\partial \bsig}$ is
\begin{eqnarray}
\frac{\partial \Phi }{\partial \bsig} = && J_{e}^{T} Q_{e}^{T} \left(Q_{e} \bfu(\bsig) - \bfd_{e}\right) \nonumber\\
  && + \beta_{e} L^{T}L (\bsig-\bsig_{\rm ref}) + {\bf y} +  {\rho} (\bsig -  p(\bom ))
\end{eqnarray}
and the search direction at each time step is:
\begin{eqnarray}
\Delta \bsig =- \left( J_{e}^{T}Q_{e}^{T}Q_{e}~J_{e} + \beta L^{T}L + \rho I \right)^{-1}
\frac{\partial \Phi }{\partial \bsig} 
\end{eqnarray}								
The model is then updated by a ``soft'' Armijo line search \cite{Nocedal06}, ${\bsig}^{k+1} = {\bsig}^{k} + \mu \Delta {\bsig}$, where the  parameter $\mu$ is adjusted to ensure sufficient decrease of $\Phi$, where $k$ refers to $k^{th}$ iteration of the geophysical descent.
We take a small fixed number of  Gauss-Newton steps (see the numerical experiment section), and as has been already mentioned, rather than searching directly for the electrical conductivity ${\bsig}$, we search for a log of conductivity ${\bf m} = \log({\bsig})$.

\bigskip
\subsection{Groundwater model block descent}
Similarly, the objective function for the augmented Lagrangian with respect to $\bom_{o}$, the initial solute mass fraction, can be written as:
\begin{eqnarray}
 \label{gwmin}
\min_{\bom_{0}}  \Phi(\bom) = &&\hf \|\bfd_{f} - Q_{f} \bom(\bom_{0})\|^{2}_{\bfSigma_{f}^{-1}} + \beta_{f}R(\bom_{0}) \nonumber \\
                              &&+{\bf y}^{\top}(\bsig -  p(\bom )) + {\frac {\rho}2} \|\bsig -  p(\bom ) \|^{2} 
\end{eqnarray}
Now the geophysical variable $\bsig$ is fixed and the derivatives of the coupling terms with respect to $\bom_{0}$ involve the sensitivity $J_{f} = \frac{\partial \bom}{\partial \bom_{0}}$, as discussed in section~\ref{GW-sens}.

Assuming a quadratic regularization for $\bom_{0}$
of the form
\begin{equation} R(\bom_{0}) = \hf\|L(\bom_{0} - \bom_{\rm ref}) \|^{2},\end{equation}
with $\bom_{\rm ref}$ being a reference solute fraction model (an initial guess or estimate from the previous step),
 the gradient of $\Phi(\bom)$ is then
\begin{eqnarray}
\frac{\partial \Phi }{\partial \bom_{0}} = &&J_{f}^{T} Q_{f}^{T} \left(\bfd_{f} - Q_{f} \bom(\bom_{0})\right) + \beta_{f} L^{T}L (\bom_{0} - \bom_{\rm ref}) \nonumber\\
&& - \eta J_{f}^{T} {\bf y} + \rho \eta J_{f}^{T} (\bsig -  p(\bom )),
\end{eqnarray}
where $p(\bom)$ represents the petrophysical constraint (Equation \eqref{minst}).
The groundwater model minimization follows the same Gauss-Newton approach as has been described above for geophysical imaging.

\subsection{ADMM stopping criteria} \label{IP-convergence}
The ADMM algorithm is stopped when the norm of the residual $r_{k}$, i.e. the constraint given by the petrophysical relationship $r_{k} = \bsig -  p(\bom )$ , is sufficiently small or the changes in $\left\|r_{k}\right\|$ between the last few iterations are bellow some threshold value. At this point, we have the capability of fitting both the geophysical and groundwater data such that the electrical conductivity and solute fraction agree.

It has been shown that ADMM has  a linear rate of convergence\footnote{Note that Gauss-Newton has a linear rate of convergence as well although the constant in Gauss-Newton may be better than the ADMM constant}, where for some applications the desired precision can be reached in a relatively small amount of steps/descents \cite{Boyd11}. The penalty term $\rho$ has an effect on the speed of convergence \cite{Ghadimi14,Deng12}, however it has been shown that under mild conditions any positive value of $\rho$ will lead to convergence \cite{Ghadimi14}, both in terms of the residual $r_{k}\rightarrow 0$ and finding an optimal solution for both models. In our study we set up $\rho$ by a trial and error procedure, where $\rho \in [0.1,1]$ provided similar results. In some cases better rate of convergence can be achieved by so called Over-relaxed ADMM, or by adding a scaling parameter for the Lagrangian multiplier, the work of \cite{Nishihara15} provides actual rate bounds based on the parameter choices.

%%%%%%%%%%%%%%%%%%%%%%%%%%%%%%%%%%%%%%%%%%%%%%%%%%%%%%%%%%%%%%%%%%%%%%%%%%%%%%%%%%%%%%%%%%%%%%%%%%%%%%%%%%%%%%%%%%%%%%%%%%%%%%%%%%%%%%%
\section{Results} 
\label{results}
\begin{figure}
\center
\begin{tabular}{cc}
\includegraphics[width=0.5\textwidth]{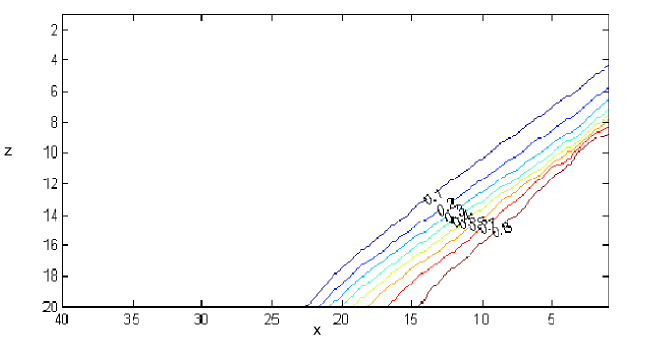} \\
\includegraphics[width=0.49\textwidth]{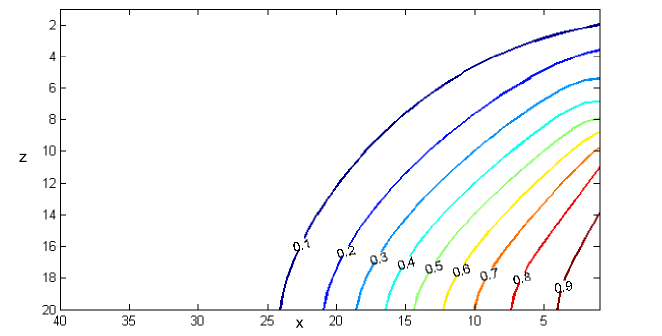}\\
\end{tabular}
\caption{\small{Contour plots of solute mass fraction for two cases of testing the Henry's benchmark problem; left: the dimensionless parameters $a = 0.33, b = 0.01$; right: $a = 0.33, b = 0.1$.}}
\label{Henry comparison}
\end{figure}

First, we tested our GW model on the Henry problem \cite{Henry64}, which is a classical benchmark problem for the VDF in 2D representing a simplified seawater intrusion case. 
We used the dimensionless parameters $a = 0.3214$ and $b = 0.1$, as in \cite{Abarca07}, and obtained similar results as in their study for a diffusive case of the Henry problem, i.e. when only molecular diffusion is consider with a fixed value for $D$ and transverse and longitudinal dispersivities being zero (see in Figure \ref{Henry comparison}). 

\subsection{Parametrization for the synthetic scenarios}
To test our method, we created different model problems in 3D representing more complex cases of seawater intrusion. We set up two cases for GW model parametrization, one with a homogeneous permeability field (Case 1) and one with heterogeneous permeability field (Case 2). The heterogeneous case is based on the field study at the Kidd2 site in the Fraser River Delta in Richmond, BC \cite{Welch01}, where the delta slope deposits confine the sandy deltaic deposits and a seawater wedge enters from the river. In both cases the boundary conditions followed the Henry's benchmark problem  with hydrostatic pressure for the seaward boundary and freshwater inflow rate for the inland boundary. The actual parameter values are presented in the Table \ref{tab-testcase} including the external fluxes representing the pumping rates.
\begin{table*} \scriptsize
\center
\caption{\small{Parametrization for the test cases.}}
\label{tab-testcase}
\begin{tabular}{c|c|c}
\hline
GW model & Heterogeneous case & Homogeneous case\\
\hline
Grid               &  $44 {\sf x} 32 {\sf x} 12$ cells  & $44 {\sf x} 32 {\sf x} 12$ cells\\
Cell size          &  $1$ x $1$ x $1$ m  & $1$ x $1$ x $1$ m\\
\hline
Permeability $\kk$ &   & \\
Silty sand         & $\kk_{x} = 2\times10^{-12} m^{2},\kk_{y} = 4.4\times10^{-11},~ \kk_{z} = 2\times10^{-14}m^{2}$  & $\kk_{x} = 4.4\times10^{-11} m^{2},$ \\
Fine and medium sand   & $ \kk_{x} = 4.4\times10^{-11} m^{2}, \kk_{y} = 4.4\times10^{-11}, \kk_{z} = 4.4\times10^{-12}m^{2}$ & $\kk_{y} = 2.4\times10^{-11},$\\
Silty clay (Fig.~\ref{fig-Kz})   & $\kk_{x} = 10^{-14} m^{2}, \kk_{y} = 4.4\times10^{-11}, \kk_{z} = 10^{-17}m^{2}$&$ ~ \kk_{z} = 1\times10^{-12}m^{2}$ \\
\hline
Porosity $\phi$    & 0.35  & 0.35\\
Dispersion $D$     & 0.0032 $m^{2}/year$ & 0.0032 $m^{2}/year$\\
Viscosity          & 0.001  & 0.001\\
\hline
Freshwater density & 1000 $kg/m^{3}$ & 1000 $kg/m^{3}$\\
Saltwater density  & 1025 $kg/m^{3}$ & 1025 $kg/m^{3}$\\
\hline
$Q_{GW}$, pumping rate up to $t_{0}$ & [x,y] = [8,14], 0.16 $d^{-1}$     & [x,y] = [8,14], 0.16 $d^{-1}$   \\
$Q_{GW}$, pumping rate up to $t_{1}$ & [x,y] = [26,26], 0.13 $d^{-1}y$    & [x,y] = [26,32], 0.13 $d^{-1}$ \\
\hline				
\end{tabular}
\vspace{0.3in}

\begin{tabular}{c|c}
\hline
Geophysical model &   \\
\hline
Grid  & 50 x 38 x 12 cells \\
Cell size & 1 to 4 m \\
\hline
Arhie's law $m$ & 1.7 \\
Background $\sigma$ &   0.0065 S/m \\
\hline
\end{tabular}
\end{table*}

\begin{figure}
\center
\includegraphics[width=0.45\textwidth]{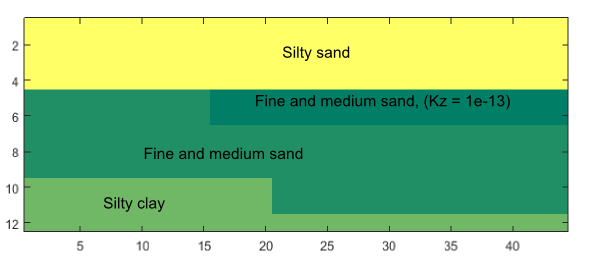}
\caption{\small{The geological layers for the heterogeneous case based on Kidd2 site in Fraser River delta \cite{Welch01}, schema for $\kk_{z}$ field.}}
\label{fig-Kz}
\end{figure}

\begin{figure*}
\centering
\begin{tabular}{cc}
 & \\
\includegraphics[width=0.5\textwidth]{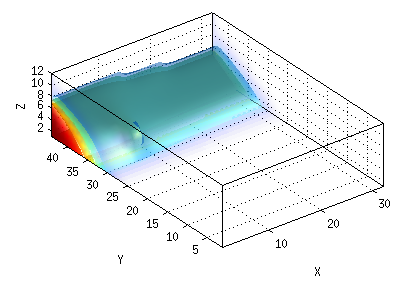} & \includegraphics[width=0.5\textwidth]{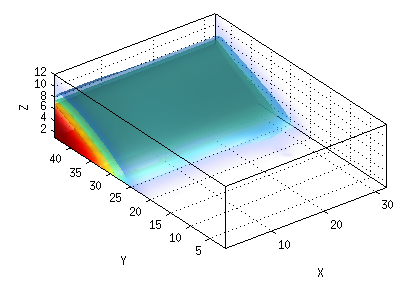}\\
\includegraphics[width=0.5\textwidth]{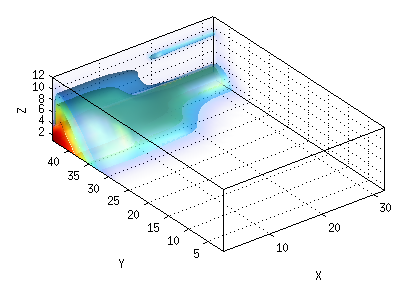} & \includegraphics[width=0.5\textwidth]{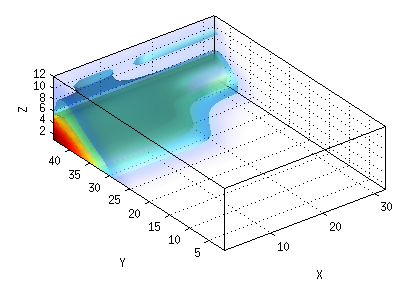}
\end{tabular}
\caption{\small{Upper left and right: Initial and final solute distribution for Case 1, Bottom left and right: Initial and final solute distribution for Case 2. Isosurfaces at $\omega$ = 0.25, 0.5 and 0.75 are plotted.}}
\label{fig_ini_final_3d}
\end{figure*}

For the initial ``unknown'' solute mass fraction distribution at time $t_{0}$ we let the GW model run forward up to a certain time. During this simulation, a pumping well is placed in the southwest part of the area.  Afterward, we altered the external fluxes, and a single pumping well was placed in the north-east area while the freshwater inflow flux was decreased.  The GW simulation then ran from the initial state at $t_{0}$ up to time $t_{1}$ for 300 days, with a time step 15 days. The ``true'' initial and final solute distributions for both cases can be seen in the Figure \ref{fig_ini_final_3d}. The external fluxes are changed at time $t_{0}$ so that the GW model, used in the coupled inversion, could not simulate the initial solute content from a zero distribution. Moreover changes to external fluxes (such as different pumping schemes or reduced discharge in the past) are also likely to happen in real conditions.

\begin{figure}
\includegraphics[width=0.5\textwidth]{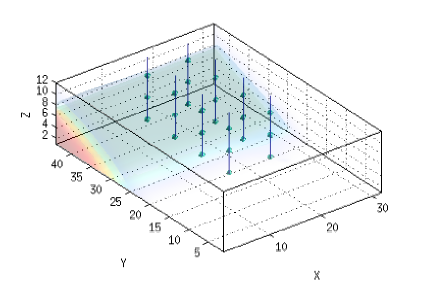}
\caption{\small{GW well sampling, 8 wells in total are placed along two transects at $x = 16~m$ and $x = 24~m$ distance, later labeled as $b$ and $c$.}}
\label{fig_GWsurvey}
\end{figure}

We collect both types of data only at time $t_{1}$. For the GW sampling we have two transects of wells (with spacing of 7 m) and 3 depth samples are collected (depth = 4,~7 and 11~m) in Case 1, and two depth samples (z = 5 and 9 m) for the heterogeneous Case 2. The position of the transects was altered for different simulations, however, here we present in detail the case with west-east locations $x = 16~m$ and $x = 24~m$, see in Figure \ref{fig_GWsurvey}. Gaussian random noise with standard deviation 0.05 was added to all measured solute fraction values. No hydraulic head or pressure GW data were used in the coupled inversion.

\begin{figure}
\includegraphics[width=0.5\textwidth]{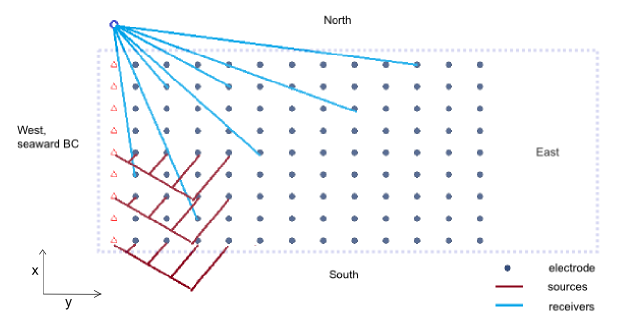}
\caption{\small{Experimental design for DC survey: Dark blue points represent the electrodes on the surface placed on a regular grid, saltwater is coming in from the west boundary.}}
\label{fig_DCsurvey}
\end{figure}

For the geophysical data, the simulated solute fraction at time $t_{1}$ was converted through Archie's law into bulk electrical conductivity, and potentials were solved through the DC forward model described in Section \ref{secGeop}. There are many different options for the electrode layout and measurement scheme, the following one was chosen based on the sensitivities of measured data, while trying to maximize the depth resolution for data collected only at the surface. The electrode layout corresponds to a regular grid with spacing 3m in $x$ direction and 4m in $y$ direction, giving in total 72 electrodes. A positive electrode was fixed close to the seaward boundary (west) and the negative charge was moving along the $x$ profile, towards east. For each source pair (72 in total) potential differences were measured on all receivers, where one of the receiver couples was always fixed and placed in the north west corner (see Figure \ref{fig_DCsurvey}). $3\%$ Gaussian random noise was added to the measured potentials.


\subsection{Coupled inversion} 
\label{res_example}
The ADMM minimization starts with the GW model descent and continues as long as the constraint residual $r_{k}$ decreases (or up to 5 runs of GW and geophysical descent). The residual $r_{k}$ represents the difference between bulk electrical conductivity and electrical conductivity derived from the GW model via Archie's law. Since this is a synthetic example we can record the actual initial and final errors next to the data misfits for both the GW and geophysical data during the minimization. By actual error we mean the norm $\epsilon(\omega_{k}) = \left\|\omega_{k} - \omega_{true}\right\|$.
Due to the ADMM approach we do not need to weight two different data misfits, however, weights still need to be assigned for the regularization term $\beta$ and the so called penalty term, $\rho$. The choice of the regularization parameter $\beta$ has already been largely discussed in the literature (see \cite{hansen,VogelBook} and reference within), and can be determined either based on initial values for $\phi_{D}, \phi_{S}$ and $\phi_{R}$, or by a trial and error procedure. The choice of the penalty parameter $\rho$ was discussed in the section \ref{IP-convergence}. A particular set of weights were applied to obtain all results presented here in; $\beta_{e} = 10^{-3},~\beta_{f} = 5\times10^{-3}$ and $\rho = 0.4.$ The number of Gauss-Newton iterations (within each ADMM descent) was 3-4 for the GW block descent, and between 6 to 10 for the geophysical block descent.

\begin{figure*}[htb]
\centering
\begin{tabular}{cc}
Case 1 & Case 2\\
\includegraphics[width=0.5\textwidth]{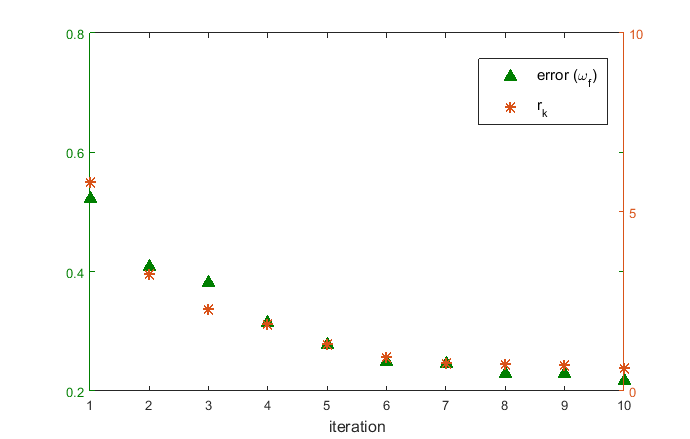}&\includegraphics[width=0.5\textwidth]{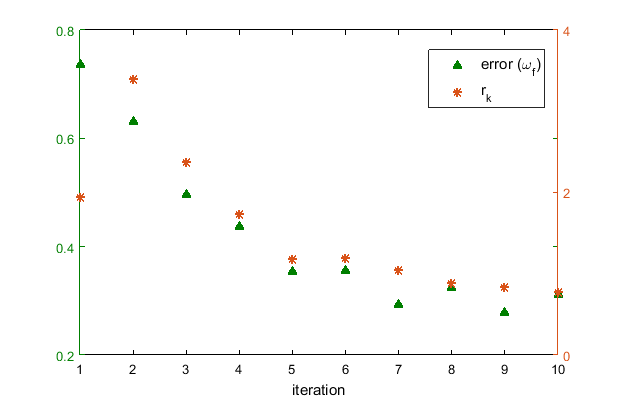}\\
\end{tabular}
\caption{\small{Green triangles represent the scaled error decrease for the final solute fraction $\bom_{f}$, orange stars correspond to updated $r_{k}$ values, where $r_{k} = \sigma_{f}(k) - p(\omega_{f}(k))$. The GW wells for this case were placed along $x = 16~m$ and $x = 24~m$}}
\label{fig_error_rk}
\end{figure*}

\begin{figure*}
\centering
\begin{tabular}{cc}
\includegraphics[width=0.5\textwidth]{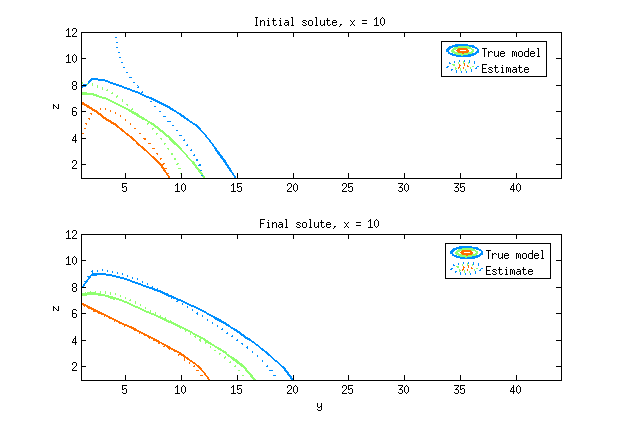} & \noindent\includegraphics[width=0.5\textwidth]{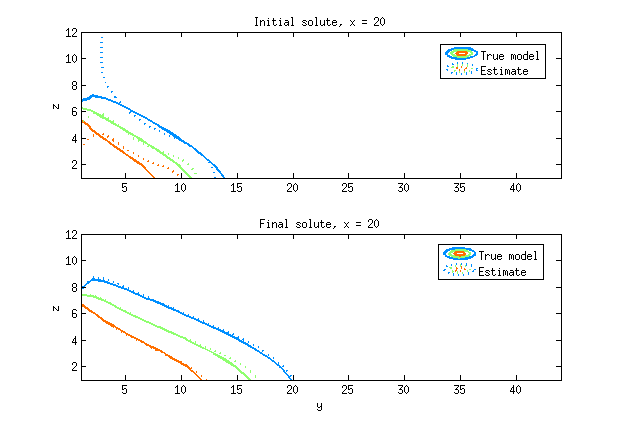}\\
\end{tabular}
\caption{\small{Case1: Contour profiles at $y = 10$ and $y = 20$. The dashed lines are estimates from the joint inversion, and the full contour lines are the actual locations corresponding to ${\boldsymbol \omega}$ = 0.25 (blue), ${\boldsymbol \omega}$ = 0.5 (green) and ${\boldsymbol \omega}$ = 0.75 (red).}}
\label{fig_profiles_homo}
\end{figure*}

\begin{figure*}[htb]
\centering
\begin{tabular}{cc}
\includegraphics[width=0.5\textwidth]{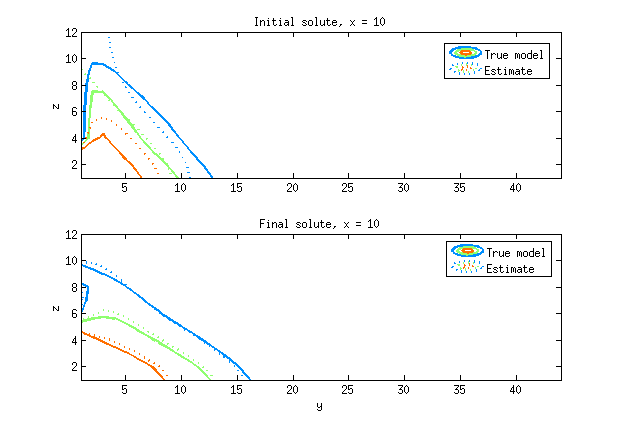} & \noindent\includegraphics[width=0.5\textwidth]{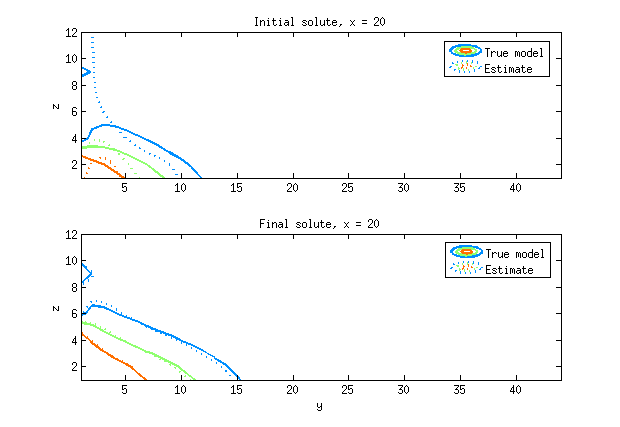}\\
\end{tabular}
\caption{\small{Case 2: Contour profiles at $x = 10$ and $x = 20$. The dashed lines are the estimates from the joint inversion, and the full contour lines are for true locations corresponding to ${\boldsymbol \omega}$ = 0.25 (blue), ${\boldsymbol \omega}$ = 0.5 (green) and ${\boldsymbol \omega}$ = 0.75(red).}}
\label{fig_profiles_hetero}
\end{figure*}

\begin{figure*}
\centering
\begin{tabular}{cc}
 & \\
\includegraphics[width=0.5\textwidth]{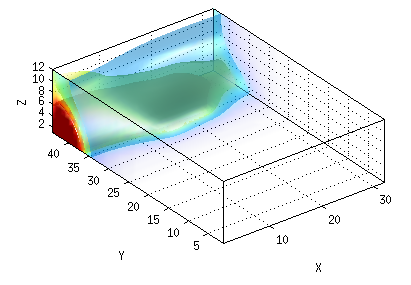} & \includegraphics[width=0.5\textwidth]{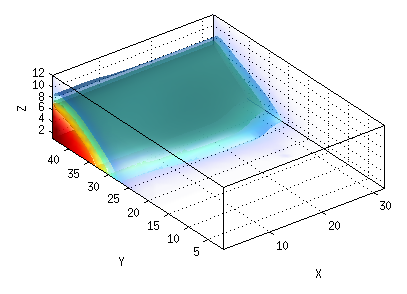}\\
\includegraphics[width=0.5\textwidth]{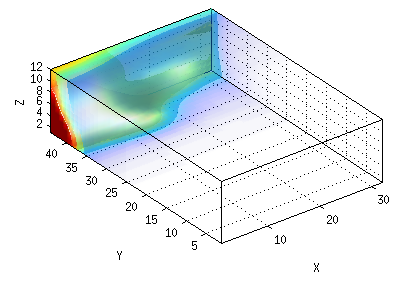} & \includegraphics[width=0.5\textwidth]{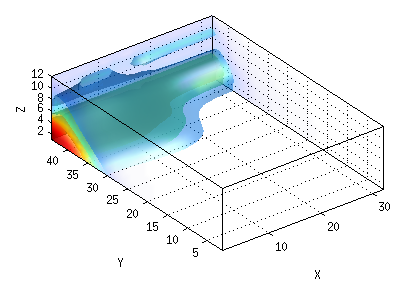}
\end{tabular}
\caption{\small{Upper left and right: Estimates for initial and final solute distribution for Case 1, Bottom left and right: Estimates for initial and final solute distribution for Case 2. Isosurfaces at $\omega$ = 0.25, 0.5 and 0.75 are plotted. The true models are plotted in Figure \ref{fig_ini_final_3d}}}
\label{est_ini_final_3d}
\end{figure*}

 In the Figure \ref{fig_error_rk}, the actual errors scaled against the error of initial estimates are plotted together with the residual $r_{k}$. Initial estimates are based on a forward simulation starting with the GW reference model. Both $r_{k}$ and $\epsilon_{k}$ decrease during the ADMM minimization. The estimates of $\omega_{0}$ can be seen in the contour profiles in Figure \ref{fig_profiles_homo}, resp. Figure \ref{fig_profiles_hetero} for $x = 10$ and $x = 20$ m or their 3D plots in Figure \ref{est_ini_final_3d}.

We also compared the ADMM with a simpler coupled approach, where both models can run more independently. First, we solve the inverse problem with GW data only and then apply Archie's law to transform the estimate to electrical conductivity at $t_{1}$. This estimate then constrains the geophysical inversion as a reference and initial model, and as such it is computationally easier to implement with no extra coupling terms in the objective function. The actual errors of solute mass fraction at $t_{0}$ and $t_{1}$ and final data misfits for the ADMM and the coupled approach are presented in Table \ref{tab-couplequ712}. For Table \ref{tab-couplequ712} we considered the lower error from GW or geophysical inversion for the coupled approach.

\begin{table}
\caption{\small{Errors, $\epsilon(\omega_{k}) = \left\|\omega_{k} - \omega_{true}\right\|$, for the solute content at time $t_{0}$ and $t_{1}$ of the two different reconstructions. Note that the ADMM provides lowest error estimation.}}% The corresponding estimate for $\bom_{f}$ are in the Figure~\ref{res3d}}}
\label{tab-couplequ712}
\begin{small}
\begin{tabular}{c|cccc}
%\begin{small}
Case 1 &   $\epsilon (\bom_{f})$ & $\epsilon (\bom_{0})$ & $\phi_{GW}(\brho \bom)$ & $\phi_{DC}(u(\bsig))$ \\
\hline
Initially                      & 14.2   &  22.6   &   974      &1046 \\
Coupled                        & 6.28   &  13.56  &   4.6      & 1.6 \\
ADMM                           & 3.1    &  10     &   1.1      & 1.4\\
\hline
\\
Case 2 &                    &                  &                                &                             \\
\hline
Initially                      & 9.3    &  16.8   &   12.4      & 1940 \\
Coupled  & 6.6    &  13.8   &   3.4      & 1.0 \\
ADMM                           & 2.8    &  11.5   &   0.8      & 0.9\\
\hline
%\end{small}
\end{tabular}
\end{small}
\end{table}

\begin{figure*}
\centering
\begin{tabular}{cc}
Case 1 & Case 2\\
\includegraphics[width=0.5\textwidth]{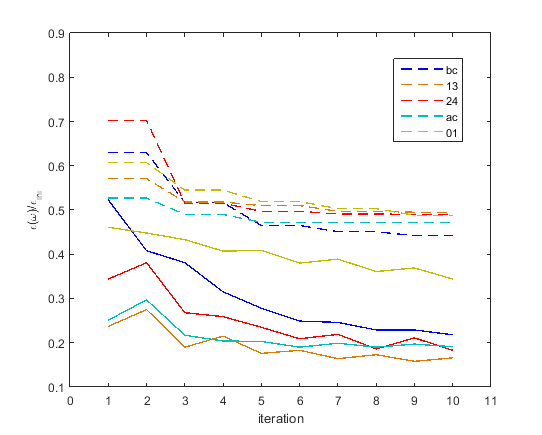} & \includegraphics[width=0.5\textwidth]{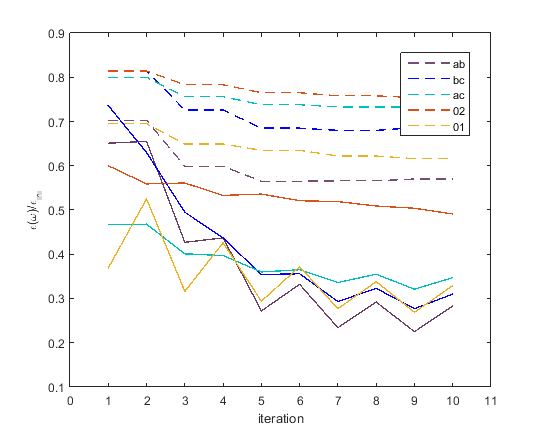}
\end{tabular}
\caption{\small{The error decrease for different GW sampling designs; the dotted line is for the $\omega_{0}$ relative error, the full line for the $\omega_{f}$ relative error decrease. In all cases the decrease slows down with further iterations. The plotted results are based on different transcet of wells plotted in Figure \ref{GWsampling}.}}
\label{all_errors1}
\end{figure*}

\begin{figure}
\includegraphics[width=0.5\textwidth]{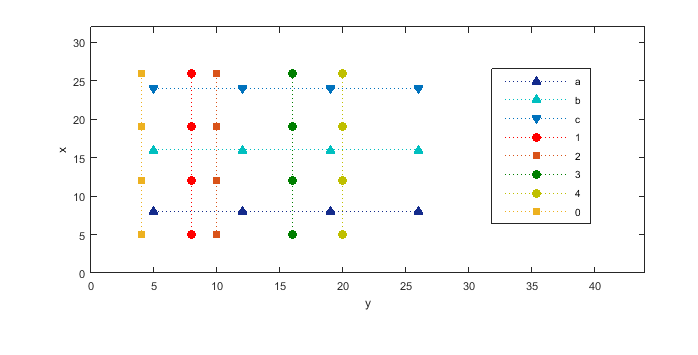}
\caption{\small{The transects along $x$ and $y$ axis in plan view. For the homogeneous Case1 the sampling depths were $z = [3,7,11]$, and for the heterogeneous Case 2 $z = [5,9]$.}}
\label{GWsampling}
\end{figure}

Additionally, we tested both methods for different locations of GW wells without changing the DC survey design. The scaled errors are plotted for all simulations in Figure \ref{all_errors1}. The different transects of wells are plotted in Figure \ref{GWsampling}. We did not use the same combinations of transects for Case 1 and 2, as the final SWI front reached further in the Case 2 compared to Case 1.

\subsection{Coupled inversion with inexact GW parameters}
\begin{figure}
\center
\includegraphics[width=0.5\textwidth]{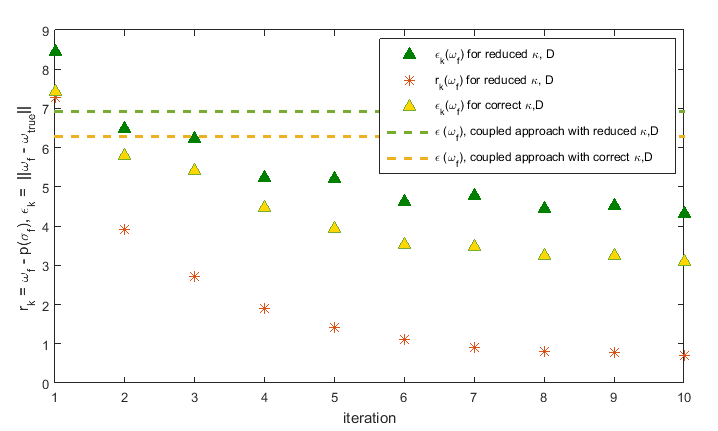}\\
\caption{\small{ Errors decrease for $\omega_{f}$ estimate and residual $r_{k}$ when correct and altered GW parameters are used in ADMM. The errors for $\omega_{f}$ for the coupled approach are plotted with a dashed line, as it is just a single value.}}
\label{Kdis70}
\end{figure}

In order to test our method for the case where the reservoir parameters are not known exactly or only approximately, we solve the problem for an inaccurate permeability field and dispersion. In the first test, we used the original GW model parameters, but decreased the homogeneous permeability field and dispersion to $70 \%$ when running the ADMM and coupled approach. The ADMM joint approach converged, but the actual errors were higher then when the correct GW parameters are used. In Figure \ref{Kdis70} you can see the error evolutions for both the ADMM with correct and incorrect GW parameters, also the error $\epsilon (\omega_{f})$ from  the coupled approach (a single value), Table \ref{tab-errors-wrongGW} provides the summary of errors for both methods.

\begin{figure}
\centering
\includegraphics[width=0.5\textwidth]{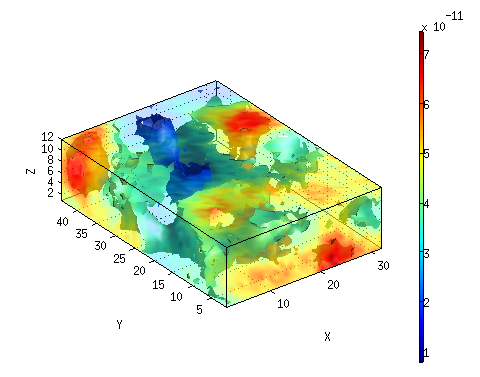}
\caption{\small{The true permeability field in $x$ direction $\kk_{x}$, when solving the inverse problem a fixed value $\kk_{x} = 4.4 \time 10^{-11}$ was used. $\kk_{y}$ and $\kk_{z}$ were also heterogeneous when creating GW data.}}
\label{Kdev3}
\end{figure}

\begin{figure}
\center
\includegraphics[width=0.5\textwidth]{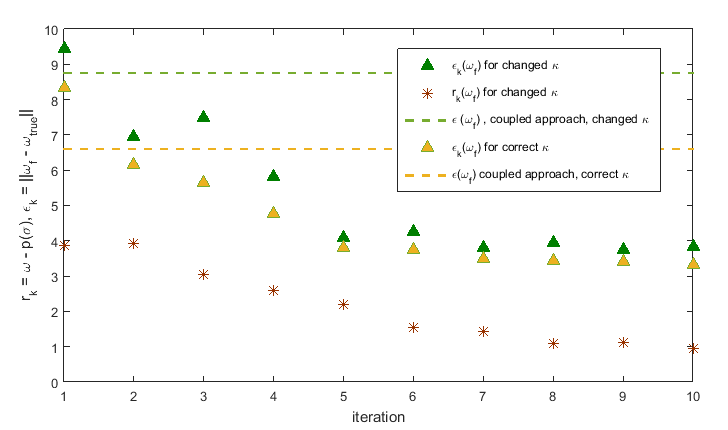}\\
\caption{\small{Errors decrease for $\omega_{f}$ estimate and residual $r_{k}$ when correct and altered GW parameters are used in ADMM. The errors for $\omega_{f}$  for the coupled approach are plotted with a dashed line.}}
\label{Kdev}
\end{figure}

In the second test, we used the homogeneous Case 1 and altered the permeability field by adding a 3D random Gaussian field to the original (see in Figure \ref{Kdev3}). The addition of the Gaussian random field thus changed the original anisotropic homogenous permeability field to a heterogenous field, which then generates a differing solute  distribution. The ratio of change in the observed solute fraction data due to different permeability compared to original data was 18$\%$. GW data based on this simulation where used in ADMM inversion, but leaving the permeability field homogeneous as in the previous calculations for Case 1. The scaled errors and residual $r_{k}$ are plotted in Figure \ref{Kdev}. For comparison we again ran the coupled approach with the same input as used in ADMM.

\begin{table}
\begin{small}
\caption{\small{Errors, $\epsilon(\omega_{k}) = \left\|\omega_{k} - \omega_{true}\right\|$, for the solute content at time $t_{0}$ and $t_{1}$ when different form true GW parameters are used in the ADMM inversion or coupled approach. Test 1 - change in permeability field, Test 2 - $70\%$ reduction in permeability and dispersion values.}}
\label{tab-errors-wrongGW}
\begin{tabular}{c|cccc}
%\begin{small}
Test 1 &   $\epsilon(\bom_{f})$ & $\epsilon(\bom_{0})$&  $\phi_{GW}(\brho \bom)$ & $\phi_{DC}(u(\bsig))$ \\
\hline
Initially                      & 12.2   &  20.7  &   63     & 1150 \\
Coupled												 & 7.35   &  16.2  &   3.64      & 2.18 \\
ADMM                           & 3.85   &  10.4  &   2.04      & 1.57\\
\hline
\\
Test 1 $\kappa$ $D$  &                    &                  &                                &                             \\
\hline
Initially                      & 17      &  22.6  &   17.2     & 1140 \\
Coupled 											 & 6.92    & 13.1   &   4.0      &  1.95\\
ADMM                           & 4.32    & 9.26   &   1.7      & 1.48\\
\hline
\end{tabular}
\end{small}
\end{table}

%%%%%%%%%%%%%%%%%%%%%%%%%%%%%%%%%%%%%%%%%%%%%%%%%%%%%%%%%%%%%%%%%%%%%%%%%%%%%%%%%%%%%%%%%%%%%%%%%%%%%%%%%%%%%%%%%
\section{Discussion}
\label{Disc}
Based on the results of the numerical examples in section 5, the ADMM approach and coupled approach proved to be advantageous compared with simple coupled aproach or separate groundwater and geophysical inversions. The splitting of the minimization procedure into the augmented Lagrangians resulted into separation of the GW and geophysical model. This separation is advantageous as the two different data mistfits do not need to be weighted in one objective function, and we can still proceed with joint minimization, which is a huge advantage. Moreover, if different geophysical survey was applied, the same ADMM method can still be applied with small adjustments to the current codes. 

To minimize each subproblem by Gauss-Newton method, we avoided using perturbation methods to calculate the sensitivities, as it is usually the case in other studies. Instead, we opted for deriving these sensitivities analytically in our codes to speed up the minimization. 
The calculation of exact sensitivities based on discretized equations allows us to work on large scale problems, avoiding expensive finite difference sensitivity calculations
and Jacobian storage. Therefore, we could perform all calculations on a standard laptop in a time scale of minutes to hours.

The ADMM  comes at a higher computation cost compared to a simple coupled approach. The additional cost is due to the repeated computation of the GW and geophysical descent, for four or five times. For comparison, using the simple coupled approach for the Case 1, we ran the forward GW model 10 times, while the ADMM required 57 forward model runs. Similarly the DC forward model required 7 runs, as opposed to 32 runs for the coupled ADMM approach.

The run time for the ADMM inversion was approximately 110 minutes for ADMM where we estimated
the relative weights. In contrast, a single coupled inversion required 27 minutes  on a standard laptop (Intel(R) Core(TM)i7-2860 QM processor and 16GB RAM). 

For comparison, when the relative weighting and regularization parameters are unknown, the ADMM method can be faster than for example the coupled Gauss-Newton optimization (substituting for $\omega_{0}$), which needs to be solved a number of times as well for different regularization parameters, however, unlike ADMM, where each iteration consists of solving a single physical model, the fully coupled
Gauss-Newton method requires solving both problems for each weighting and regularization parameters.

Note also that 3D inversions are well known to be computationally intensive, where stochastic approaches, or minimization codes with the sensitivities calculated using perturbation methods, can require time frame of weeks \cite{Irving10,Commer14}. In contrast, our approach of computing the sensitivities allows us to solve 3D problems in minutes. 

The joint inversion with ADMM achieved a lower error for both the initial and final solute fraction distributions in all tested examples compared to simple coupled approach. In the example presented above in detail in section \ref{res_example} the error for $\omega_{f}$ estimate by ADMM was roughly $50\%$ of the error by a coupled approach in homogeneous case, and $60\%$ for the heterogeneous case. For the initial solute fraction $\omega_{0}$ the improvement by ADMM was  $60\%$ and $70\%$ respectively. 

The estimates for the final solute distribution were generally better than for the initial solute mass fraction due to the fact that the DC data were collected at this time, and also the coupling constraint between $\bom_{f}$, solute fraction, and $\bsig_{f}$, electrical conductivity was enforced for the final time $t_{1}$. In all cases the ADMM converged to minimum, though it shows some of its typical aspects: a relatively quick drop during the first few iterations and a slow decrease towards the end. However, for the case of groundwater modeling applications, the initial decrease in error might be sufficient. Our synthetic study confirmed the theoretical results about the convergence of this method.

The goodness of estimates with low errors is largely determined by the quality and amount of data available, which play a key role in success of solving any inverse problem. It is apparent from Figure \ref{all_errors1} that different data sampling locations result in differing final model errors, even though the ADMM method followed the same pattern of error decrease. Due to the inverse problem setup, where the other GW model parameters and boundary conditions are known, increasing the amount of GW data led to little improvement of the ADMM approach compared to the coupled approach. In contrast if the amount of GW data samples was reduced, the ADMM method was still able to give reasonable estimates with lower errors. The experimental design and locations of GW samples are crucial when solving the inverse problem. For example, if some of the intruding seawater wedges are not captured by GW wells, the success of the joint approach is entirely dependent on the geophysical data.  
The uncertainty and credibility of both groundwater and geophysical data also implies there is uncertainty in the derived estimates of the hydrological states, which will remain regardless of the method applied to solve the inverse problem. The coupled approach has the advantage of revealing possible discrepancies between the two sources of information.

For the first set of simulations with Case 1 and Case 2 we assumed the GW model parameters were known, excluding initial and actual solute content. This can be regarded as an overly simplifying approach, but is justified for testing the feasibility of the joint inversion strategy. To demonstrate the robustness of the ADMM method with respect to variations in GW parameter values, we altered the GW model parameters in the inversion process. As expected, this led to estimates with higher error compared to solving the problem with the correct parameters. Nevertheless, the ADMM converged to estimates with lower error than the simple coupled approach (see in Figure \ref{Kdis70}), as it could be partially ``corrected'' by information from geophysical data. We are aware that further increasing of the error in the GW model parametrization would also lead to worse estimates, but in that case any coupled approach will not be able to provide a more accurate estimates.

 The ADMM method can only be applied to constrained optimization problems. For the hydrogeophysical applications this implies that the petrophysical relationship constraint should have a low uncertainty as it is ``enforced'' during the minimization. Therefore, in environments where we expect varying electrical conductivity of geological material it would not be a recommended approach. If we still want to follow the methods described above, the coupling term based on the petrophysical relationship would stay as a part of the objective function, but not as a constraint, and a block coordinate descent method could be used for minimization. However, the sensitivities derived for ADMM approach would not change. An alternate option would be to use similarity measures based on gradient fields supporting a similar structure of two different models, without forcing a perfect matching relationship. Some of the proposed methods will be subject of future research.

\section{Summary} \label{sum}
In this paper we have developed a hydrogeophysical inversion framework which improves estimates of hydrological states by jointly inverting both sources of data. To alleviate computational costs in the inverse problem, sensitivities were analytically derived as opposed to using perturbation methods. An alternating direction method of multipliers was applied to  keep the codes for each inverse subproblem relatively separate, with only a few changes necessary to proceed with a joint inversion. The ADMM approach presented here enabled us to invert both the geophysical and groundwater data at once. 

Based on the synthetic cases tested, the new approach improves the estimates of the initial and current solute content when compared with a simple coupled or direct substitution approach. Moreover, the results from the synthetic experiments suggest that even if the GW parameters used in the inversion deviate from the true parameter set, the ADMM still manages to converge while improving the estimates of solute fraction for both the initial and final time. Not surprisingly, the efficiency and gain of the ADMM joint inversion compared to inverting GW data alone is dependent on the quality of both data sets.

%\begin{acknowledgements}
%\end{acknowledgements}

\end{document}